# Direct tensor processing with coherent light

Yufeng Zhang,[1,2,3][†] Xiaobing Liu,[4,5][†] Chenguang Yang,[1] Jinlong Xiang,[1,3] Hao Yan,[1,6] Tianjiao Fu,[4,5] Kaizhi Wang,[1,6]* Yikai Su,[1,3]* Zhipei Sun,[2]* Xuhan Guo[1,3]*

[1.] School of Electronic Information and Electrical Engineering, Shanghai Jiao Tong University, Shanghai 200240, China.

[2.] Department of Electronics and Nanoengineering, Aalto University, Espoo 02150, Finland.

[3.] State Key Laboratory of Advanced Optical Communication Systems and Networks, Shanghai Jiao Tong University, Shanghai 200240, China.

[4.] Changchun Institute of Optics, Fine Mechanics and Physics, Chinese Academy of Sciences, Changchun 130033, China.

[5.] University of Chinese Academy of Sciences, Beijing 100049, China.

[6.] Yiwu Zhiyuan Research Center of Electronic Technology, Jinhua 322000, China.

Corresponding authors: Kaizhi Wang, Yikai Su, Zhipei Sun, Xuhan Guo

[†]These authors contributed equally to this work.

**Abstract:**

Tensor processing is the cornerstone of modern technological advancements, powering critical applications in data analytics and artificial intelligence. While optical computing offers exceptional advantages in bandwidth, parallelism, and energy efficiency, existing methods optimized for scalar operations struggle to efficiently handle tensor-based tasks, limiting their applicability in complex applications, such as neural networks. Here, we report Parallel Optical Matrix-Matrix Multiplication (POMMM), a novel paradigm that enables fully parallel tensor processing through a single coherent light propagation. This approach addresses key limitations of current optical methods, scaling the performance with data dimension, while improving theoretical computational power and efficiency. We demonstrate its high consistency with GPU-based matrix–matrix multiplication across both real-valued and complex-valued domains. Moreover, we showcase its adaptability, scalability, and versatility in tensor processing applications such as convolutional and vision transformer neural networks. Furthermore, we analyse the theoretical compatibility and efficiency of POMMM in relation to existing optical computing paradigms, highlighting its potential to outperform current state-of-the-art methods. By enabling a variety of computational tasks and supporting multi-wavelength and large-scale expansion, POMMM provides a scalable, high-efficient foundation for advancing next-generation optical computing.

**Introduction**

Artificial intelligence (AI) technology, driven by neural networks, is rapidly reshaping various aspects of modern life. However, the training and inference processes of these networks impose significant computational demands on digital processors. A key operation in these computations is tensor processing, typically implemented through matrix-matrix multiplication (MMM) on GPU tensor cores[1-4]. This approach, however, poses significant challenges, including large memory bandwidth demands, substantial power consumption, and inefficiently using tensor core resources. In contrast, optical computing, with its inherent characteristics of large bandwidth, high parallelism, and low energy consumption, has emerged as a promising platform to compute and accelerate neural networks[5,6]. State-of-the-art optical computing paradigms enable parallel dot product[7,8], vector-matrix multiplication[9-14], diffraction computing[15,16], and Fourier transform[17-20] by modulating the amplitude, phase, and wavelength of light during a single propagation through planar waveguides or free space, laying the basis for various optical neural network (ONN) implementations[21-26].

While optical vector-matrix multiplication-driven ONNs are capable of executing tensor processing, they typically require multiple light propagations for each MMM operation, which limits their parallelism and overall efficiency. Some approaches have attempted to improve parallelism by leveraging space multiplexing[27,28], wavelength-time multiplexing[29,30], wavelength-space multiplexing[31-34] or pre-computed inverse operations[21,35]. However, these methods are constrained by device[28,36] and system design[29,37] and, in some cases, even contradict the fundamental principle of optical parallel acceleration at the computational logic level[38,39]. In contrast, while diffraction-based computing offers high parallelism, it relies on diffraction principles to construct the computing architecture[40-42], making it unsuitable for general-purpose computation. As a result, existing optical computing paradigms struggle to balance both high parallelism and generality. They are often tailored to specific types of neural network computations or individual processing steps, rather than providing a universal acceleration framework for various complex neural network steps within a single system. This limitation also hinders the seamless deployment of advanced network models developed for GPUs onto optical computing platforms in the future.

Here, we report a novel direct Parallel Optical Matrix-Matrix Multiplication (POMMM) paradigm for tensor processing that operates through a single coherent light propagation. This approach leverages the bosonic properties of light[43] and exploits the duality between the spatial position, phase gradients and the spatial frequency distribution to enhance optical computing capabilities, adding an extra computational dimension without relying on additional time, space, wavelength multiplexing or pre-processing. We validate this paradigm through theoretical simulations and the construction of a physical optical prototype, demonstrating its strong consistency with standard GPU-based MMM over various input matrix scales. Building upon the POMMM simulation and prototype as fundamental computational units, we develop a GPU-compatible ONN framework and demonstrate the direct optical deployment of different GPU-based neural network architectures such as convolutional neural networks (CNNs)[44] and vision transformer (ViT) networks[45,46], incorporating various tensor operations such as multi-channel convolution, multi-head self-attention, and multi-sample fully connected layer[47]. We explored the scalability of POMMM, including its data type, multi-wavelength multiplexing for high dimensional tensor processing, and large-scale computing ability for complex neural networks and tasks. Finally, through a comprehensive comparison with existing optical computing paradigms, we highlight the superior performance of the POMMM paradigm as a transformative approach alongside the ONN framework it enables, offering theoretically improved efficiency and versatility in tensor processing. Our results demonstrate that POMMM has the potential to achieve more complex and higher-dimensional general parallel optical computing, enabling future computing demands.

**Results**

Conventionally, MMM between matrix A and matrix B is executed in two sequential steps. First, the Hadamard product (element-wise multiplication) is performed between each row of matrix A and each column of matrix B. Subsequently, the results of these Hadamard products are summed and assigned to their corresponding elements in the output matrix. Typically, the optical Hadamard product can be implemented by modulating the amplitude of spatial light, while the optical Fourier transform, which enables approximate summation or integration, can be achieved through quadratic phase modulation (e.g., passing through a lens)[48] (Fig. 1a, left panel). However, a key challenge remains: how to compute the dot products (sum after Hadamard product) of all rows of matrix A with all columns of matrix B (or all rows of matrix $B^T$, transpose of matrix B) in parallel and map them to their corresponding positions without mutual interference.

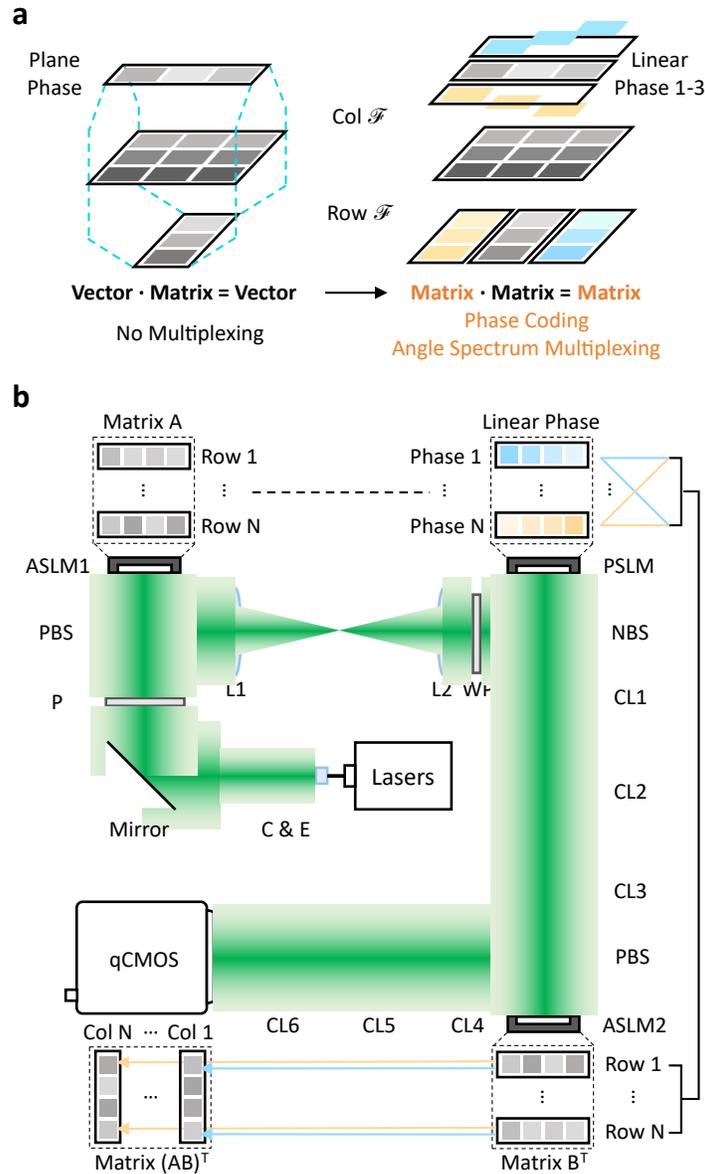

**Fig. 1 Principle and experimental setup of POMMM. a** Basic principle of POMMM compared with Stanford method. $\mathscr{F}$, Fourier transform. **b** Experimental setup and corresponding matrix operations. C & E, Collimation

and Expansion; P, Polarizer; PBS, Polarization Beam Splitter; ASLM, Amplitude Spatial Light Modulator; L, Lens; WP, Wave Plate; NPBS, Non-Polarization Beam Splitter; PSLM, Phase Spatial Light Modulator; CL, Cylindrical Lens; qCMOS, quantitative Complementary Metal Oxide Semiconductor camera.

Fortunately, the Fourier transform exhibits two well-known properties: the time-shifting property and the frequency-shifting property. The former states that shifting a signal in the time (space) domain does not alter the amplitude distribution of its time (space) spectrum, while the latter indicates that applying a linear phase modulation in the time (space) domain results in a corresponding frequency shift in the time (space) spectrum. Leveraging these two properties, we construct the core concept of POMMM (Fig. 1a, right panel). First, the elements of matrix A are encoded onto the amplitude and position of a spatial optical field, with each row modulated by a linear phase with special gradient. Then, a column-wise Fourier transform is applied to the optical field carrying the complex amplitude signal, which can be implemented by imaging along the row direction and focusing along the column direction. At this stage, each row of the optical field represents the superposition of all rows of matrix A. Subsequently, amplitude modulation is applied to perform the Hadamard product between this "hybrid matrix" and the matrix $B^T$, effectively enabling the parallel computation of all row-column Hadamard products. Finally, a row-wise Fourier transform is performed to complete the summation. Due to the presence of different linear phase modulations in the row dimension, the contributions from different rows of matrix A naturally separate into distinct positions in the final computational result (details in Supplementary Note 1).

To experimentally validate POMMM using conventional optical components and demonstrate its functionality and compatibility, we design an optical proof-of-concept prototype (Fig. 1b). We first encode matrix A onto a collimated beam using an amplitude spatial light modulator (ASLM1). This encoded beam is then imaged onto a phase spatial light modulator (PSLM) through a 4f optical system, where distinct linear phase modulations are applied to each row (this step can also be realized by complex amplitude modulators). Subsequently, the beam passes through a cylindrical lens (CL1, 2, 3) assembly: two CLs (1, 3) acting along the row direction and one CL (2) along the column direction. This setup enables row-wise imaging and column-wise Fourier transform at the surface of the second amplitude spatial light modulator (ASLM2), effectively superimposing the "fan-out" results from each row of matrix A in space. ASLM2 then encode matrix $B^T$ onto this combined field via amplitude modulation. Finally, the beam then passes through another CL assembly (4, 5, 6) performing the inverse functionality, achieving row-wise Fourier transform and column-wise imaging, and the output matrix $(AB)^T$ is captured by a qCMOS camera (details in Methods and Supplementary Note 2). Importantly, this entire tensor operation is fully parallel, with single-shot generating all values simultaneously.

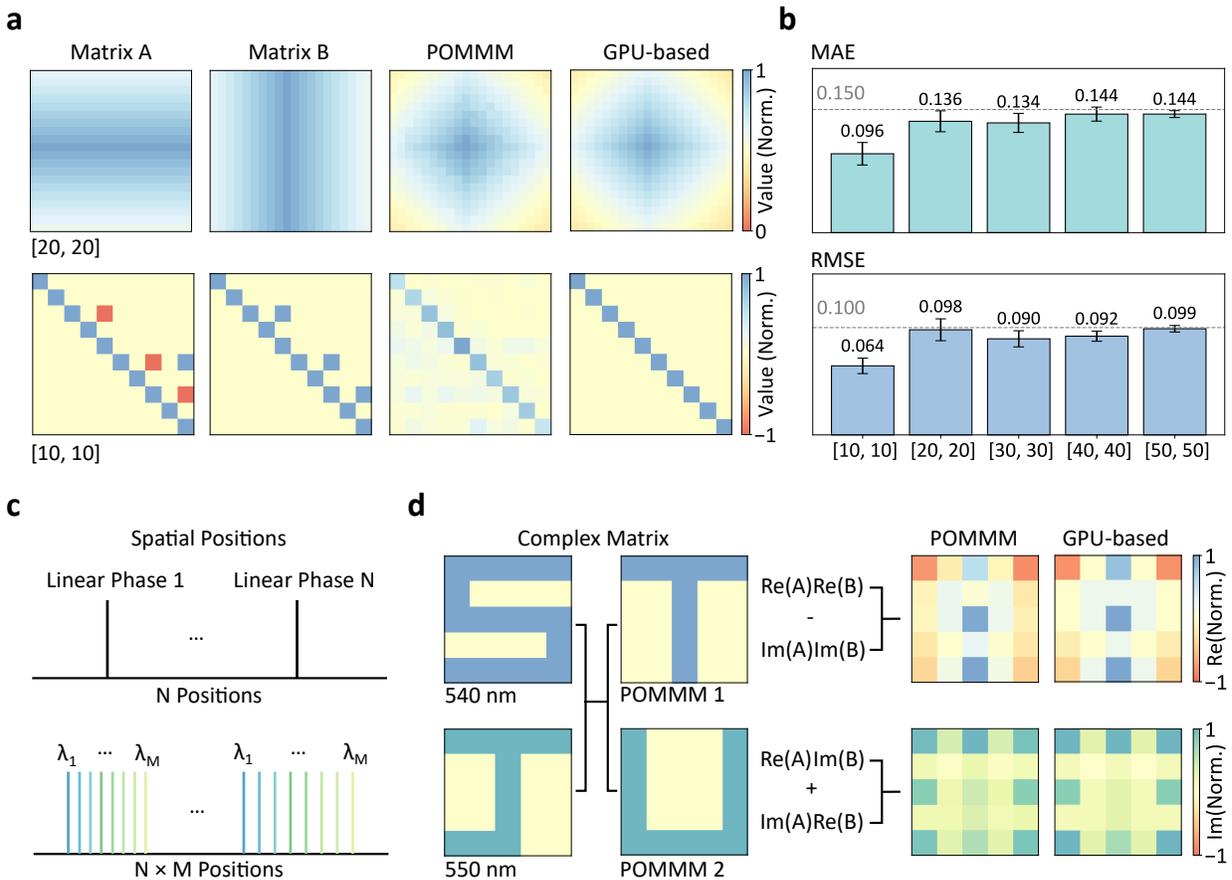

**Fig. 2 Demonstration and evaluation of POMMM. a** Experimental POMMM results compared with GPU-based results. Upper panel, non-negative matrix multiplication. Bottom panel, real matrix multiplication (original light field in Fig. S5, S6). **b** Comparison between experimental results of POMMM and GPU-based MMM on 50 random matrices of varying sizes. Upper panel, Mean Absolute Error (MAE). Bottom panel, normalized Root Mean Square Error (RMSE). **c** Paradigm of wavelength multiplexing extension of POMMM. **d** Demonstration of wavelength

multiplexing POMMM by two complex matrices. "S" and "J" are the real and imaginary parts of complex matrix A, respectively, which are modulated to 540 nm and 550 nm wavelengths. After two wavelength multiplexing POMMMs, they respectively perform tensor-matrix multiplication with the real part "T" and imaginary part "U" of complex matrix B to obtain a complete complex result matrix (original light field in Fig. S9).

To validate the reliability of the POMMM paradigm, we compared its experimental results with those obtained from GPU-based MMM across various scenarios. These included non-negative matrices of different sizes, such as symmetric and upper-triangular matrices (Fig. 2a, Fig. S5a), as well as real-valued matrices including conjugate matrix pairs (Fig. 2a, Fig. S6a). All comparisons demonstrated strong consistency with the GPU results. Furthermore, we conducted a large-sample quantitative analysis across multiple matrix sizes ([10, 10], [20, 20], [30, 30], [40, 40], [50, 50], with 50 random matrix pairs for each size) and evaluated the computational errors between POMMM and GPU-based MMM (Fig. 2b). The results showed that both the mean absolute error (less than 0.15) and the normalized root-mean-square error (less than 0.1) remained low, confirming the accuracy and reliability of the POMMM framework. In addition, we performed a detailed analysis of the theoretical sources of error in POMMM, along with the experimental prototype's imperfections (Fig. S3, S7, S8). Based on this, we proposed and verified effective error-suppression strategies through theoretical simulations (details in Supplementary Note 3). Due to its reliance on single-shot coherent optical propagation, each linear phase modulation combined with a Fourier transform naturally functions as an independent spectrometer. This inherent property makes multi-wavelength extension in POMMM straightforward. The results for different wavelengths can be spatially separated and simultaneously obtained in a single-shot measurement (Fig. 2c). Leveraging this, we encoded the real and imaginary parts of a complex-valued matrix (with patterns "S" and "J") onto two distinct wavelengths and performed two rounds of multi-wavelength POMMM with another complex matrix (patterns "T" and "U") to realize full complex MMM. The results show a high degree of consistency with those computed by GPU (Fig. 2d), and this process can be equivalent to the real-valued multiplication between a [2, 5, 5] tensor and two [5, 5] matrices. In summary, our experimental demonstrations confirm that POMMM can reliably perform MMM across varying scales and data types using single-shot optical propagation. Moreover, its inherent compatibility with multi-wavelength extensions suggests strong potential for scaling to single-shot tensor-matrix multiplication (N⁴ parallelism).

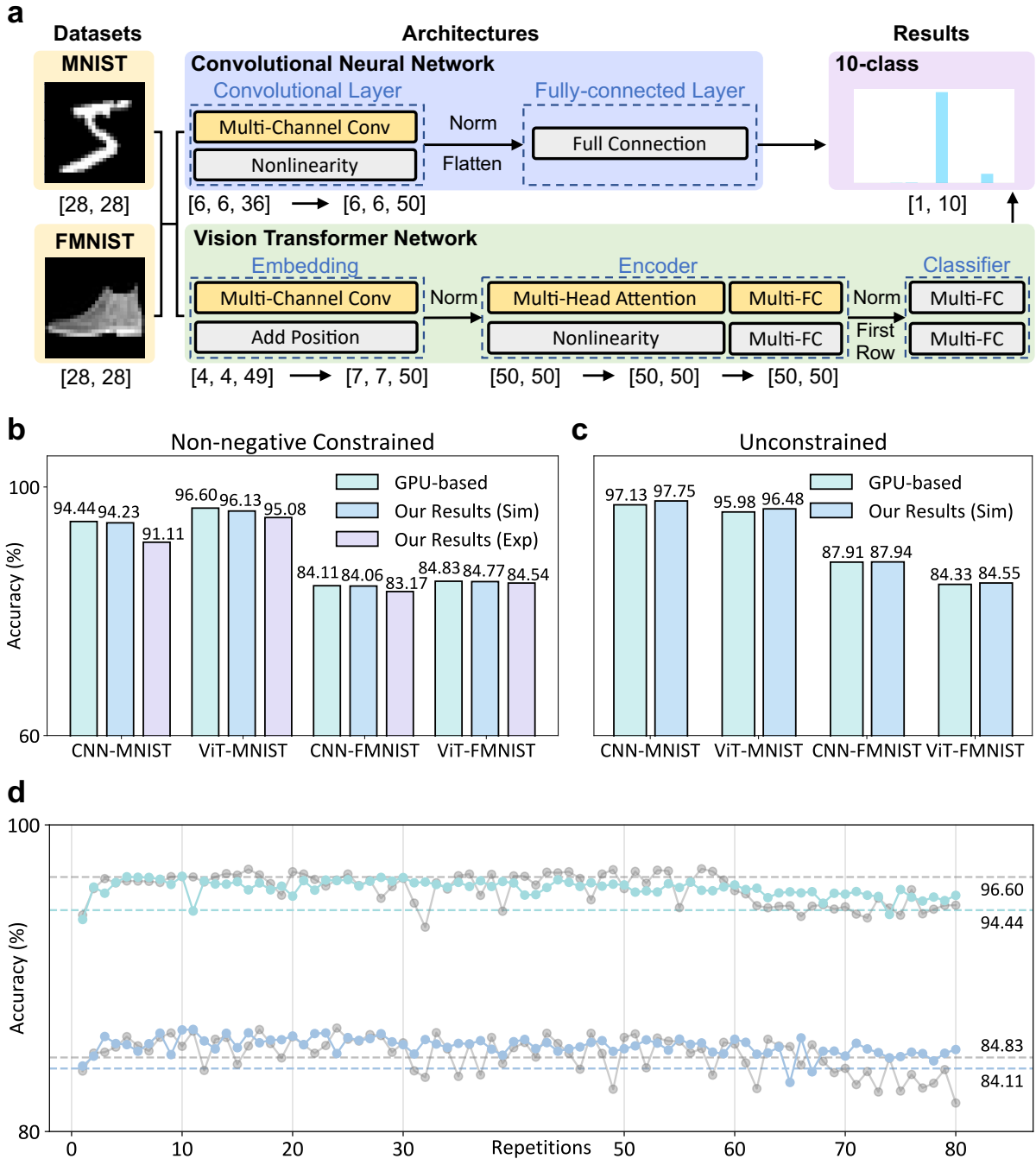

**Fig. 3 Applications of POMMM for GPU-compatible tensor processing in different neural networks. a** Entire process of CNN (blue area) and ViT network (green area). CNN includes a convolution layer and a fully-connected

layer, while ViT network includes an embedding layer, an encoder, and a classifier. Where the yellow steps are POMMM-implemented, and the rest are GPU-implemented. **b** Tests of direct deployment of non-negative constrained GPU-trained weights across different datasets on POMMM simulation and prototype (confusion matrices are shown in Fig. S13). **c** Tests of direct deployment of all-step unconstrained GPU-trained weights across different datasets on POMMM simulation (confusion matrices are shown in Fig. S16). **d**. Comparison of ONN inference accuracy based on POMMM with different computing errors (related to element repetitions). The colored curves are CNN models, the grey curves are ViT models, and the dashed lines are the corresponding baseline of GPU-based non-negative models.

Due to the high consistency between POMMM and GPU tensor core-based MMM, POMMM theoretically enables the direct deployment of standard GPU-based neural network architectures. This eliminates the need to design custom network architectures tailored to the unique optical propagation constraints of conventional ONN approaches. To further validate this capability, we conducted direct inference experiments using both CNN and ViT networks on our POMMM simulation and prototype, utilizing MNIST[49] and Fashion-MNIST[50] (FMNIST) datasets (Fig. 3a). These architectures include three representative tensor processing steps common in modern neural networks: multi-channel convolution, multi-head self-attention, and multi-sample fully connected layers (details in Methods). We first used four different models (CNN-MNIST, ViT-MNIST, CNN-FMNIST, and ViT-FMNIST) trained on GPUs with non-negative weights and tested inference across GPU, simulated POMMM, and POMMM prototype (details in Supplementary Note 4, and experimental results in Fig. S11 and S12). The inference outputs across all platforms showed high agreement, demonstrating that POMMM supports various tensor processing operations and allows for the direct deployment of GPU-trained weights (Fig. 3b). Furthermore, we employed two successive simulated POMMM units to directly execute all linear operations with unconstrained GPU-trained weights (Fig. S14, S15), achieving inference accuracy highly consistent with GPU results (Fig. 3c). Additionally, we explored the scalability of POMMM by performing an image style transfer task[51] based on a U-Net model[52], where the largest MMM reached a scale of [256, 9216] × [9216, 256] (Fig. S17). These findings highlight the versatility and scalability of the POMMM framework, confirming its potential to theoretically support all linear tensor computations in diverse neural networks.

Moreover, in scenarios where the error between POMMM and GPU-based MMM is non-negligible (e.g., spectral leakage due to low repetitions, Fig. S7), POMMM can still be employed by model training. We evaluated inference performance across 80 different error levels for each network architecture and task, where models trained directly using the POMMM kernel achieved inference accuracy comparable to the GPU reference (Fig. 3d). This indicates that strict consistency between POMMM and GPU-based MMM is not always necessary. With appropriate training, high-quality inference remains achievable, and recent advances in onsite training for ONNs support the practical realization of such approaches[53,54], thereby reducing future engineering requirements for POMMM deployment. Interestingly, when nonlinear activation functions are disabled (e.g., ReLU: $f(x) = max(0, x)$, is inactive due to non-negative constraints, leading CNNs to degenerate into linear classifiers), models trained with POMMM under certain error levels outperformed their GPU-based counterparts (colored curve in Fig. 3d). This suggests that under linear constraints, POMMM may enhance model expressiveness after training compared to standard linear classifiers.

In conclusion, through both theoretical simulations and experimental validation, we have demonstrated that POMMM can directly deploy GPU-trained neural network models. We have also explored its potential for deep, large-scale, and nonlinear extensions, revealing the promising applicability of POMMM in future tensor processing tasks.

**Discussion**

Compared with existing general-purpose optical computing paradigms, POMMM demonstrates a significant theoretical computational advantage under both single-wavelength and multi-wavelength extensions (Fig. 4a). Furthermore, the experimentally demonstrated computing scale confirms the practical scalability of POMMM for real-world applications (Fig. S18). In our experiments, the prototype was entirely constructed using off-the-shelf, non-specialized components, resulting in a practical energy efficiency of only 2.62 GOP/J. However, since POMMM requires only passive phase modulation (excluding data input and output), it is, in principle, compatible with a wide range of free-space optical computing devices[8]. It's extremely high theoretical computing dimensionality enables a dramatic increase in effective performance when integrated with high-speed, large-scale, and dedicated photonic hardware (Fig. 4b), making POMMM an ideal computational paradigm for next-generation optical computing platforms. Nevertheless, compared with dedicated paradigms such as diffractive and scattering-based computing that leverage complex-valued operations, POMMM introduces additional phase modulation to enable real-valued operations. This may increase deployment and cascading complexity in practice, potentially limiting its convenience and performance in vision-related tasks (details in Supplementary Note 5).

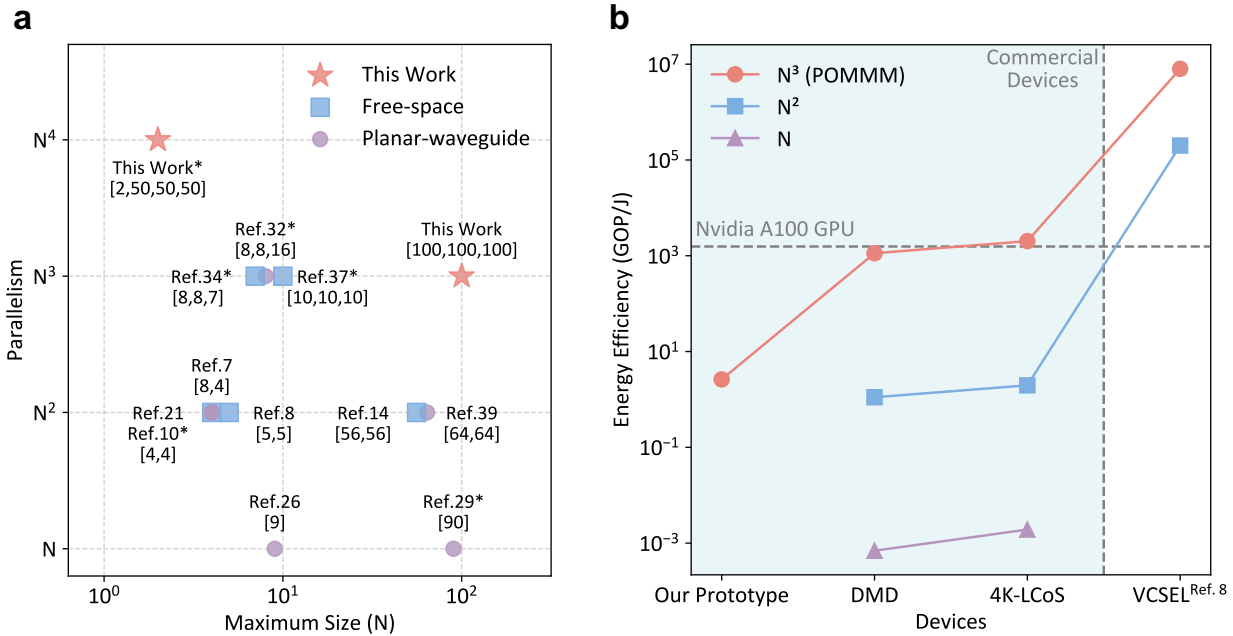

**Fig. 4 Performance analysis of POMMM. a** Theoretical computing power compared with existing optical computing paradigms (Table. S1). * multi-wavelength multiplexing. The vertical axis represents the single-shot computational parallelism, while the horizontal axis indicates the actual computational scale achieved in experiments (denoted by the smaller dimension). **b** Practical energy efficiency evaluation. DMD, digital micromirror device; LCoS, liquid crystal on silicon; VCSEL, vertical-cavity surface-emitting laser. The vertical axis denotes the energy efficiency, and the horizontal axis represents different device platforms. Each curve corresponds to a computational paradigm with a specific level of theoretical computing power. The data point on the far left indicates the actual performance of our prototype (in Table S2), while all other values are estimated under ideal conditions (i.e., full device utilization, in Table S3).

## Methods

**Simulation and experimental details:**

The theoretical simulations and experimental demonstrations of POMMM are conducted following the workflow outlined in Fig. 1b. The simulation framework incorporates three distinct propagation models (Fast Fourier Transform (FFT), Huygens–Fresnel theory, and Rayleigh–Sommerfeld theory) and accounts for critical system factors including device parameters, aperture effects, discrete periodicity, and diffraction limitations. Extensive simulations under the paraxial approximation demonstrate high consistency among these models (Fig. S7, S8). Unless otherwise specified, all simulated optical field distributions presented in this work are based on the Rayleigh–Sommerfeld theory, whereas the ONN training and inference simulations employ the FFT method.

For the POMMM experimental demonstrations (Fig. 1b), a 532 nm solid-state continuous-wave laser is used as the light source for single-wavelength tests, while a broadband pulsed laser (400-2200 nm) in conjunction with a multi-wavelength tunable filter (430-1450 nm) is used for multi-wavelength tests. All laser beams are collimated and expanded, then polarized using a linear polarizer (Thorlabs, LPVISE100-A) to obtain p-polarized light. The beam subsequently passes through a PBS (Thorlabs, PBS251) and is directed onto the surface of ASLM1 (Upolabs, HDSLM80R-A, 8 μm pixel pitch, 1200 × 1920 resolution, 8-bit depth, 60 Hz frame rate, 95% zero-order diffraction efficiency). ASLM1 encodes matrix A via amplitude modulation while converting the p-polarization to s-polarization, which is reflected by PBS. The resulting optical field is relayed to the surface of a PSLM (Upolabs, HDSLM80R-P Pro, identical pixel pitch and resolution, 10-bit depth) using a 4f imaging system composed of two achromatic lenses (Thorlabs, ACT508-200-A, f = 200 mm). The beam also passes through a half-wave plate (Lbtek, MAHWP20-VIS) and a 50:50 NBS (Thorlabs, BS013), converting s-polarization back to p-polarization. The PSLM imposes phase modulation on the beam, which is then reshaped by a cylindrical lens assembly. This system includes two cylindrical lenses (Thorlabs, LJ1567L-1-A, f = 100 mm) for the row dimension and one cylindrical lens (Thorlabs, LJ1653L2-A, f = 200 mm) for the column dimension. The modulated beam is then incident on ASLM2 (identical to ASLM1), which encodes matrix $B^T$ in amplitude and converts p-polarization to s-polarization, reflected by PBS2 (Thorlabs, PBS251). Finally, the beam passes through a second cylindrical lens group (two cylindrical lenses (Thorlabs, LJ1567L-1-A, f = 100 mm) for the column dimension and one cylindrical lens (Thorlabs, LJ1653L1-A, f = 200 mm) for the row dimension) before being captured by a high-resolution qCMOS camera (Hamamatsu, ORCA-Quest qCMOS C15550-20UP, 16-bit depth). Detailed descriptions of the experimental procedures for the POMMM demonstration are provided in Supplementary Note 2.

**Optical CNN structure:**

The CNN framework consists of a multi-channel convolutional layer with 50 channels, followed by a ReLU activation function and a standard fully connected layer. The input images from the MNIST and Fashion-MNIST datasets are initially sized at [28, 28]. After applying padding of [1, 2, 1, 2], the image dimensions become [31, 31]. With a convolution kernel size of [6, 6] and a stride of 5, the convolution operation requires $6 \times 6 = 36$ dot products between a [6, 6] matrix slice of the input and the [6, 6] kernel to cover the entire [31, 31] matrix. This results in a convolution output of size [6, 6]. For multi-channel convolution, this process transforms a [6, 6, 36] tensor into a [6, 6, 50] tensor. The 36 dot products between 36 [6, 6] matrix slices and a [6, 6] kernels can be interpreted as a vector-matrix multiplication between a [36, 36] matrix and a [36, 1] vector. Thus, tensor processing in the 50-channel convolution can be abstracted as a MMM between a [36, 36] matrix and a [36, 50] matrix (Fig. S10a). The result is then flattened into a [1, 1800] vector and passed through a fully connected layer, producing a [1, 10] vector. This operation can be abstracted as a vector-matrix multiplication between a [1, 1800] vector and a [1800, 10] matrix. The resulting [1, 10] vector represents the ten-class classification output, which is processed using a Softmax function. The loss is computed using the cross-entropy loss function with the image label vector. Backward propagation is then applied to update the gradients and complete the training.

**Optical ViT network structure:**

The ViT network consists of an embedding layer, an encoder, and a classifier. In the embedding layer, patch embeddings are generated using a 50-channel convolution, like the multi-channel convolution in the CNN. The difference lies in the convolution kernel size, which is [4, 4], and the stride, which is 4. This results in 7 × 7 = 49 dot products between the [4, 4] kernel and the [28, 28] input matrix to cover the image. This tensor processing can also be abstracted as an MMM between a [49, 16] matrix and a [16, 50] matrix, producing a [49, 50] matrix. A [1, 50] class token is then appended to the patch embeddings, resulting in a [50, 50] matrix. In the encoder layer, multi-head self-attention is applied, where a query (Q) is mapped to a set of key (K) - value (V) pairs: attention = $QK^TV$. Q, K, and V are obtained by performing MMM with the input matrix of size [50, 50] and the weight matrices of $W_Q$, $W_K$, and $W_V$, each of size [50, 50]. The input to the encoder layer is added to the output of the self-attention and passed through two multiple fully connected layers to produce a [50, 50] matrix as the encoder's output. The tensor processing in the multi-fully connected layer can be abstracted as an MMM between a [50, 50] matrix and a [50, 50] matrix (Fig. S10b). The classifier layer extracts the [1, 50] class token from the normalized output of the encoder layer. This class token is then passed through two simple fully connected layers, resulting in a [1, 10] vector representing the classification output.

## Data Availability

All data are available in the main text and the supplementary information.

## Code Availability

The code for POMMM simulation and ONN test can be found at the following website: https://github.com/DecadeBin/POMMM.git

## References

1       Cichocki, A. *et al.* Tensor decompositions for signal processing applications: From two-way to multiway component analysis. *IEEE Signal Processing Magazine* **32**, 145-163 (2015).
2       LeCun, Y., Bengio, Y. & Hinton, G. Deep learning. *Nature* **521**, 436-444 (2015).
3       Shastri, B. J. *et al.* Photonics for artificial intelligence and neuromorphic computing. *Nature Photonics* **15**, 102-114 (2021).
4       Choquette, J., Gandhi, W., Giroux, O., Stam, N. & Krashinsky, R. Nvidia a100 tensor core gpu: Performance and innovation. *IEEE Micro* **41**, 29-35 (2021).
5       Caulfield, H. J. & Dolev, S. Why future supercomputing requires optics. *Nature Photonics* **4**, 261-263 (2010).
6       Wetzstein, G. *et al.* Inference in artificial intelligence with deep optics and photonics. *Nature* **588**, 39-47 (2020).

7　Zuo, Y. *et al.* All-optical neural network with nonlinear activation functions. *Optica* **6**, 1132-1137 (2019).

8　Chen, Z. *et al.* Deep learning with coherent VCSEL neural networks. *Nature Photonics* **17**, 723-730 (2023).

9　Reck, M., Zeilinger, A., Bernstein, H. J. & Bertani, P. Experimental realization of any discrete unitary operator. *Physical Review Letters* **73**, 58 (1994).

10　Yang, L., Ji, R., Zhang, L., Ding, J. & Xu, Q. On-chip CMOS-compatible optical signal processor. *Optics Express* **20**, 13560-13565 (2012).

11　Clements, W. R., Humphreys, P. C., Metcalf, B. J., Kolthammer, W. S. & Walmsley, I. A. Optimal design for universal multiport interferometers. *Optica* **3**, 1460-1465 (2016).

12　Goodman, J. W., Dias, A. & Woody, L. Fully parallel, high-speed incoherent optical method for performing discrete Fourier transforms. *Optics Letters* **2**, 1-3 (1978).

13　Farhat, N. H., Psaltis, D., Prata, A. & Paek, E. Optical implementation of the Hopfield model. *Applied Optics* **24**, 1469-1475 (1985).

14　Spall, J., Guo, X., Barrett, T. D. & Lvovsky, A. Fully reconfigurable coherent optical vector–matrix multiplication. *Optics Letters* **45**, 5752-5755 (2020).

15　Lin, X. *et al.* All-optical machine learning using diffractive deep neural networks. *Science* **361**, 1004-1008 (2018).

16　Liu, C. *et al.* A programmable diffractive deep neural network based on a digital-coding metasurface array. *Nature Electronics* **5**, 113-122 (2022).

17　Chang, J., Sitzmann, V., Dun, X., Heidrich, W. & Wetzstein, G. Hybrid optical-electronic convolutional neural networks with optimized diffractive optics for image classification. *Scientific reports* **8**, 1-10 (2018).

18　Bueno, J. *et al.* Reinforcement learning in a large-scale photonic recurrent neural network. *Optica* **5**, 756-760 (2018).

19　Miscuglio, M. *et al.* Massively parallel amplitude-only Fourier neural network. *Optica* **7**, 1812-1819 (2020).

20　Hu, Z. *et al.* High-throughput multichannel parallelized diffraction convolutional neural network accelerator. *Laser & Photonics Reviews* **16**, 2200213 (2022).

21　Shen, Y. *et al.* Deep learning with coherent nanophotonic circuits. *Nature Photonics* **11**, 441-446 (2017).

22　Spall, J., Guo, X. & Lvovsky, A. I. Hybrid training of optical neural networks. *Optica* **9**, 803-811 (2022).

23　Moralis-Pegios, M., Giamougiannis, G., Tsakyridis, A., Lazovsky, D. & Pleros, N. Perfect linear optics using silicon photonics. *Nature Communications* **15**, 5468 (2024).

24　Pintus, P. *et al.* Integrated non-reciprocal magneto-optics with ultra-high endurance for photonic in-memory computing. *Nature Photonics*, 1-9 (2024).

25　Tsakyridis, A. *et al.* Photonic neural networks and optics-informed deep learning fundamentals. *APL Photonics* **9** (2024).

26　Xu, S. *et al.* Optical coherent dot-product chip for sophisticated deep learning regression. *Light: Science & Applications* **10**, 221 (2021).

27    Ma, G. *et al.* Dammann gratings-based truly parallel optical matrix multiplication accelerator. *Optics Letters* **48**, 2301-2304 (2023).

28    Ma, G., Yu, J., Zhu, R. & Zhou, C. Optical multi-imaging–casting accelerator for fully parallel universal convolution computing. *Photonics Research* **11**, 299-312 (2023).

29    Xu, X. *et al.* 11 TOPS photonic convolutional accelerator for optical neural networks. *Nature* **589**, 44-51 (2021).

30    Xu, S., Wang, J., Yi, S. & Zou, W. High-order tensor flow processing using integrated photonic circuits. *Nature Communications* **13**, 7970 (2022).

31    Yeh, P. & Chiou, A. E. Optical matrix–vector multiplication through four-wave mixing in photorefractive media. *Optics Letters* **12**, 138-140 (1987).

32    Feldmann, J. *et al.* Parallel convolutional processing using an integrated photonic tensor core. *Nature* **589**, 52-58 (2021).

33    Dong, B. *et al.* Partial coherence enhances parallelized photonic computing. *Nature* **632**, 55-62 (2024).

34    Luan, C., Davis III, R., Chen, Z., Englund, D. & Hamerly, R. Single-Shot Matrix-Matrix Multiplication Optical Tensor Processor for Deep Learning. *arXiv preprint arXiv:2503.24356* (2025).

35    Jiao, L. *et al.* AI meets physics: a comprehensive survey. *Artificial Intelligence Review* **57**, 256 (2024).

36    Fan, Y. *et al.* Dispersion-assisted high-dimensional photodetector. *Nature* **630**, 77-83 (2024).

37    Latifpour, M. H., Park, B. J., Yamamoto, Y. & Suh, M.-G. Hyperspectral in-memory computing with optical frequency combs and programmable optical memories. *Optica* **11**, 932-939 (2024).

38    Chen, Y. 4f-type optical system for matrix multiplication. *Optical Engineering* **32**, 77-79 (1993).

39    Hua, S. *et al.* An integrated large-scale photonic accelerator with ultralow latency. *Nature* **640**, 361-367 (2025).

40    Wang, T. *et al.* An optical neural network using less than 1 photon per multiplication. *Nature Communications* **13**, 123 (2022).

41    Fu, T. *et al.* Photonic machine learning with on-chip diffractive optics. *Nature Communications* **14**, 70 (2023).

42    Chen, Y. *et al.* All-analog photoelectronic chip for high-speed vision tasks. *Nature* **623**, 48-57 (2023).

43    Bose. Plancks gesetz und lichtquantenhypothese. *Zeitschrift für Physik* **26**, 178-181 (1924).

44    Krizhevsky, A., Sutskever, I. & Hinton, G. E. Imagenet classification with deep convolutional neural networks. *Advances in neural information processing systems* **25** (2012).

45    Vaswani, A. Attention is all you need. *Advances in Neural Information Processing Systems* (2017).

46    Dosovitskiy, A. An image is worth 16x16 words: Transformers for image recognition at scale. *arXiv preprint arXiv:2010.11929* (2020).

47    LeCun, Y., Bottou, L., Bengio, Y. & Haffner, P. Gradient-based learning applied to document recognition. *Proceedings of the IEEE* **86**, 2278-2324 (1998).

48    Goodman, J. W. *Introduction to Fourier optics*.   (Roberts and Company publishers, 2005).

49    Deng, L. The mnist database of handwritten digit images for machine learning research. *IEEE signal processing magazine* **29**, 141-142 (2012).

50  Xiao, H., Rasul, K. & Vollgraf, R. Fashion-mnist: a novel image dataset for benchmarking machine learning algorithms. *arXiv preprint arXiv:1708.07747* (2017).
51  Simonyan, K. Very deep convolutional networks for large-scale image recognition. *arXiv preprint arXiv:1409.1556* (2014).
52  Ronneberger, O., Fischer, P. & Brox, T. in *Medical image computing and computer-assisted intervention–MICCAI 2015: 18th international conference, Munich, Germany, October 5-9, 2015, proceedings, part III 18.*   234-241 (Springer).
53  Xue, Z. *et al.* Fully forward mode training for optical neural networks. *Nature* **632**, 280-286 (2024).
54  Spall, J., Guo, X. & Lvovsky, A. I. Training neural networks with end-to-end optical backpropagation. *Advanced Photonics* **7**, 016004-016004 (2025).

## Contributions

Conceptualization: Y.Z., K.W., Y.S., Z.S., X.G.; Methodology: Y.Z., X.L.; Investigation: Y.Z., K.W., X.L., J.X.; Visualization: Y.Z., Z.S., X.G., H.Y.; Supervision: K.W., X.G., Y.S., Z.S.; Validation: Y.Z., X.L., C.Y.; Software: X.L., Y.Z.; Writing—original draft: Y.Z., Z.S., X.G., H.Y., J.X., X.L.; Writing—review & editing: Y.Z., X.L., Z.S., X.G.

## Acknowledgements

This work is supported by the National Key Research and Development Program of China under contract nos. 2023YFB2804702, in part by the Natural Science Foundation of China under contract nos. 62175151, 62071297, and 62341508, and in part by Shanghai Municipal Science and Technology Major Project under contract nos. BH030071. We gratefully acknowledge the Shanghai Institute of Optics and Fine Mechanics, Chinese Academy of Sciences, for providing access to a multi-wavelength tunable filter for our experiments.

## Competing interests

The authors declare that they have no competing interests.